\documentclass[12pt]{article}
\usepackage{graphicx,amsmath,amssymb,color,bm}
\usepackage{ulem}
\usepackage[numbers,sort&compress]{natbib}

\begin{document}

\bigskip

\bigskip
\centerline {\Large\textbf {Novel Magnetic Quantization of
Bismuthene}}
%\centerline {\Large\textbf {Novel Magnetic Quantization of sp$^{3}$ Bonding in
%Monolayer Tin}}

%\centerline {\Large \textbf {}}\vskip0.6 truecm

\centerline{Szu-Chao Chen$^{1,\ast}$, Jhao-Ying Wu$^{2,\star}$, and Ming-Fa
Lin$^{3,\dagger}$ }
\centerline{$^{1}$Center for Micro/Nano Science and
Technology, National Cheng Kung University, Tainan, Taiwan 701}
\centerline{$^{2}$Center of General Studies, National Kaohsiung Marine University, Kaohsiung, Taiwan 811}
\centerline{$^{3}$Department of Physics, National Cheng Kung University,
Tainan, Taiwan 701}\vskip0.6 truecm

\noindent The generalized tight-binding model, being based on the
spin-dependent sublattices, is developed to explore the magnetic
quantization of monolayer bismuthene. The sp$^{3}$ orbital hybridizations,
site energies, nearest and next-nearest hopping integrals, spin-orbital
interactions and magnetic field (${B_{z}}$ ${\hat{z}}$) are taken into
account simultaneously. There exist three groups of low-lying Landau levels
(LLs), in which they are mainly from the (6p$_{x}$,6p$_{y}$,6p$_{z}$) orbitals,
and only the first group belongs to the unoccupied conduction states.
Furthermore, each group is further split into the spin-up- and
spin-down-dominated subgroups. The six subgroups present the rich and unique
$B_{z}$-dependent LL energy spectra, covering the specific or arc-shaped $%
B_{z}$-dependences, the normal/irregular spin-split energies, and the
non-crossing/crossing/anti-crossing behaviors. Specially, the second group
of valence LLs near the Fermi level can create the frequent inter-subgroup
LL anti-crossings since the main and side modes are comparable. The main
features of energy spectra can create the special structures in density of
states.

\vskip0.6 truecm

$\mathit{PACS}$:73.22.-f,73.20.At

\newpage

\bigskip The monoelemental 2D materials have stirred a lot of experimental
and theoretical researches since the first discovery of graphene in 2004 by
the mechanical exfoliation \cite{1}. They are very suitable for studying the
diverse physical, chemical and material properties. Up to now, the
successfully synthesized group-IV and group-V systems cover
few-layer/multilayer graphene \cite{1,2,3}, silicene \cite{Si1,Si2},
germanene \cite{Ge2,Ge3}, tinene \cite{Sn1}, phosphorene \cite{P1,P3},
antimonene \cite{Sb1,Sb2}, and bismuthene \cite{Bi1,Bi2,Bi3,Bi4,Bi5,Bi6}.
Specifically, layered bismuthene are epitaxially grown on the 3D Bi$_{2}$Te$%
_{3}$(111)/Bi$_{2}$Se$_{3}$(111)/Si(111) substrates \cite%
{Bi1,Bi2,Bi3,Bi4,Bi5}. Also they could be directly obtained from the
mechanical exfoliation \cite{Bi6}. There are various Hamiltonians, being
sensitive to the planar/buckled structures, lattice symmetries, layer
numbers, stacking configurations, single- or multi-orbital hybridizations,
site energies, spin-orbital couplings (SOCs), and external electric and
magnetic fields. How to solve new Hamiltonians is one of the main-stream
topics in physics science. This work is focused on the unique quantization
of monolayer bismuthene in a uniform perpendicular magnetic field (${B_{z}}$
${\hat{z}}$) by using the generalized tight-binding model.

For few-layer bismuthene, there are some studies on geometric structures
\cite{Bi3,Bigeo,Bigeo1,Bigeo2,Bigeo3}, electronic structures \cite%
{Bigeo,Bigeo1,Bigeo2,Bigeo3} and transport properties \cite{Bi6}. Bi atoms
could form 2D honeycomb lattice with a highly buckled structure, as clearly
identified from the measurements of scanning tunneling microscopy and
high-resolution electron diffraction \cite{Bi3,Bi4,Bigeo2}. The
angle-resolved photoemission spectroscopy has confirmed the unusual
low-lying energy bands with the cone-like, parabolic, and sombrero-shaped
dispersions centered at the $\Gamma $ point \cite{Bi1,Bi2,Bi3,Bi4,Bi5}. The
similar band structures are revealed in the theoretical calculations using
the first-principles method \cite{Bigeo,Bigeo1,Bigeo2,Bigeo3} and the
tight-binding model \cite{Bigeo1}. They are deduced to be dominated by the
geometric structure, the multi-orbital chemical bondings, the distinct site
energies and the strong SOC. Such critical factors are responsible for a 3D
rhombohedral Bi semimetal \cite{Bibulk,Bibulk1}. The combination with a
magnetic field is expected to create the diversified Landau levels (LLs) in
terms of the $B_{z}$-dependent energy spectra and the quantum oscillation
modes of the spatial probability distributions. This is worthy of a
systematic investigation.

In general, there are two theoretical methods in studying the
magneto-electronic properties. The effective-mass model is to make a
perturbation expansion about the high-symmetry point and then do the
magnetic quantization. It is suitable and reliable for layered systems with
simple and monotonous band structures, such as, the rich $B_{z}$-dependent
LL energy spectra in monolayer graphene \cite{effgraphene,effgraphene1},
silicene, germanene \cite{effsige}, MoS$_{2}$ \cite{effMoS2}, and few-layer
AA- and AB-stacked graphenes \cite{AAAB,AAAB1,AAAB2}. This method will
become cumbersome for condensed-matter systems in the presence of
complicated interactions and composite/non-uniform external fields. As for
the generalized tight-binding model, all the significant interactions and
the various external fields are included in the calculations simultaneously.
For example, it has predicted three kinds of LLs in sliding bilayer graphene
with various stacking configurations \cite{slidinggraphene}, and the unusual
features of LLs in ABC- and AAB-stacked graphene \cite{ABCAAB,ABCAAB1}. On
the experimental side, scanning tunneling spectroscopy (STS) is the most
powerful instrument in measuring the magneto-electronic energy spectra,
since the differential tunneling conductance is roughly proportional to
density of states (DOS). Up to now, STS measurements are successful in
identifying the magnetically quantized energies in AB-stacked graphite \cite%
{STSgraphite,STSgraphite1} and few-layer AB-stacked graphene \cite%
{STSgraphene,STSgraphene1}.

We utilize the generalized tight-binding model to investigate
magneto-electronic properties of monolayer bismuthene. The magnetic
Hamiltonian matrix elements, being related to the sp$^{3}$ bondings $\&$
site energies, SOC and $B_{z}$ $\widehat{z}$, are calculated from the
tight-binding functions in an enlarged unit cell due to vector potential.
The orbital-, spin- and sublattice-decomposed wave functions are delicately
evaluated to characterize the dominating oscillation modes and determine the
quantum numbers of LLs. The dependences of the LL energies and wave
functions on the field strength are explored in detail, especially for the
neighboring LL energy spacings, the spin-split energies, and the
non-crossing/crossing/anti-crossing behaviors. The LL anti-crossings will be
examined from the probability transfer between the main and side modes, as
well as the change of spin configurations. The van Hove singularities in
creating the special structures of DOS are also discussed.

Bismuthene is composed of buckled hexagonal lattices in which two equivalent
A and B sublattices are, respectively, located at two parallel planes with a
separation of $\Delta _{z}$=$1.81$ \AA . The primitive unit vectors are
indicated by $\mathbf{a}_{1}$ and $\mathbf{a}_{2}$ with a lattice constant
of a=$4.33$ \AA\ (Fig. 1(a)), and the buckled structure is associated with
the angle between the Bi-Bi bond and the $z$-axis, $\theta $=$126^{\circ }$
(Fig. 1(b)). The generalized tight-binding model \cite{GTBM,GTBM1} is
utilized to explore the electronic properties under external fields, in
which all the critical interactions are taken into account simultaneously.
The strong sp$^{3}$ orbital hybridizations, the distinct site energies and
the significant SOC will dominate the essential properties near the Fermi
level. In the bases of \{$\left\vert 6p_{z}^{A}\right\rangle ,\left\vert
6p_{x}^{A}\right\rangle ,\left\vert 6p_{y}^{A}\right\rangle ,\left\vert
6s^{A}\right\rangle ,\left\vert 6p_{z}^{B}\right\rangle ,\left\vert
6p_{x}^{B}\right\rangle ,\left\vert 6p_{y}^{B}\right\rangle ,\left\vert
6s^{B}\right\rangle $\}$\otimes \left\{ \uparrow ,\downarrow \right\} ,$ the
Hamiltonian is expressed as

\begin{eqnarray}
H &=&\underset{i,o,m}{\sum }E_{o}C_{iom}^{+}C_{iom}^{{}}+\underset{%
\left\langle i,j\right\rangle ,o,o^{\prime },m}{\sum }\gamma _{oo^{\prime
}}^{\mathbf{R}_{ij}}C_{iom}^{+}C_{jo^{\prime }m}^{{}}+\underset{\left\langle
\left\langle i,j\right\rangle \right\rangle ,o,o^{\prime },m}{\sum }\gamma
_{oo^{\prime }}^{\mathbf{R}_{ij}}C_{iom}^{+}C_{jo^{\prime }m}^{{}}  \notag \\
&&+\underset{i,p_{\alpha },p_{\beta }^{{}},m,m^{\prime }}{\sum }\frac{%
\lambda _{\text{SOC}}}{2}C_{ip_{\alpha }m}^{+}C_{ip_{\beta }m^{\prime
}}^{{}}(-i\epsilon _{\alpha \beta \gamma }\sigma _{mm^{\prime }}^{\gamma }),
\end{eqnarray}%
$\newline
$where $C_{iom}^{+}$($C_{iom}^{{}}$), $i$, $o$, and $m$ stand for the
creation (annihilation) operator, lattice site, atomic orbital, and spin,
respectively. The first \ term is the site energy, and E$_{o}$ of the 6s and
6p orbitals are set to be $-9.643$ eV and $-0.263$ eV, respectively \cite%
{Bibulk1}. The second term is the nearest-neighbor hopping integral ($\gamma
_{oo^{\prime }}^{\mathbf{R}_{ij}}$) which depends on the type of atomic
orbitals, the translation vector of the nearest-neighbor atom ($\mathbf{R}%
_{ij}$), and $\theta $. The various interactions are characterized by the sp$%
^{3}$ chemical bondings: ${V_{pp\pi }=-0.679}$ eV, ${V_{pp\sigma }=2.271}$
eV, ${V_{sp\sigma }=1.3}$ eV and ${V_{ss\sigma }=-0.703}$ eV \cite{Bibulk1},
as clearly indicated in Fig. 1(c). For the next-nearest-neighbor atoms, the
hopping integral in the third term is independent of $\theta $, since such
atoms are located at the same plane. It is related to ${V_{pp\pi }=0.004}$
eV, ${V_{pp\sigma }=0.303}$ eV, ${V_{sp\sigma }=0.065}$ eV and ${V_{ss\sigma
}=-0.007}$ eV. The last term represents the intra-atomic SOC $V_{\text{soc}}$%
=$\lambda _{\text{soc}}\overrightarrow{L}\cdot \overrightarrow{s}$ with ${%
\lambda _{\text{soc}}\,=1.5}$ eV. $\alpha ,\beta $ and $\gamma $ denote the $%
x$, $y$ or $z$ direction, and $\sigma $ is the Pauli spin matrix. $V_{\text{%
soc}}$ can also be expressed as $V_{\text{soc}}$=$\lambda _{\text{soc}}(%
\frac{L_{+}s_{-}+L_{-}s_{+}}{2}+L_{z}s_{z}),$ where $L_{\pm }$ and $s_{\pm }$
are the ladder operators for the angular momentum and spin. The SOC is
vanishing for the same orbital. It could induce the change of spin
configurations between the 6p$_{z}$ and (6p$_{x}$,6p$_{y}$) orbitals $\&$
the 6p$_{x}$ and 6p$_{y}$ orbitals.

The periodical Peierls phases, being created by a uniform perpendicular
magnetic field, can modulate the hopping integral as\ $\gamma _{oo^{\prime
}}^{\mathbf{R}_{ij}}(\mathbf{B}_{z})$=$\gamma _{oo^{\prime }}^{\mathbf{R}%
_{ij}}\exp i(\frac{2\pi }{\Phi _{0}}\int_{r_{j}}^{r_{i}}\mathbf{A}(\mathbf{r}%
)\cdot d\mathbf{r})$ and induce an enlarged unit cell. $\Phi _{0}$ ($hc/e$)
is the flux quantum. Under the Landau gauge $\mathbf{A}$=$(0,B_{z}x,0)$, a
rectangular unit cell covers $4R_{B}$ ($4\times 25500/B_{z}$) Bi atoms (Fig.
1(a)), where $R_{B}$ is the ratio of $\Phi _{0}$ versus magnetic flux
through a hexagon. The area of a reduced Brillouin zone (a small rectangle
in Fig. 1(d)) is $4\pi ^{2}/\sqrt{3}a^{2}R_{B}$. The magnetic Hamiltonian is
built from the space spanned by the $32R_{B}$ tight-binding functions \{$%
\left\vert A_{om}^{i}\right\rangle ;\left\vert B_{om}^{i}\right\rangle $\},
where $i=1,2,...;2R_{B}$. By the detailed analytic calculations, this
Hermitian matrix could be transferred into the band-like form to solve LL
energies and wave functions more efficiently. When a uniform electric field
is applied along the $z$-axis, it can create a Coulomb potential $V_{z}/2$ ($%
-V_{z}/2$) on the site energy of the A (B) sublattice. The generalized
tight-binding model could be further developed to comprehend the magnetic
quantization in other layered systems with complex orbital bondings and spin
configurations under composite fields.

Monolayer bismuthene exhibits a feature-rich electronic structure due to the
significant multi-orbital bondings, site energies and SOC. The low-lying
electronic properties are mainly determined by three energy bands near the $%
\Gamma $ point (Fig. 1(d)), as clearly shown in Fig. 2. Each electronic
state is doubly degenerate for the spin degree of freedom. The first
conduction ($c_{1}$) has parabolic energy dispersion centered at the $\Gamma
$ point. The second valence band presents the valley-like dispersion except
a slightly rounded structure near the $\Gamma $ point. Specially, the first
valence band ($v_{1}$) reveals the sombrero-shaped structure with the
non-monotonic wavevector dependence, in which the extreme points deviate
from the $\Gamma $ point. There coexist two constant-energy loops within a
certain energy range, being expected to induce the complicated magnetic
quantization under their strong competitions. Apparently, the lowest
unoccupied state in the $c_{1}$ band and the highest occupied states in the $%
v_{1}$ band (the outer constant-energy loop) have the different wave vectors
and thus lead to an indirect energy gap of $0.293$ eV.

Bismuthene exhibits three groups of low-lying LLs with distinct
characteristics, as clearly shown in Figs. 3 and 4. The valence and
conduction LLs are asymmetric about the Fermi level. The first, the second,
and the third groups are associated with the magnetic quantization of the $%
c_{1}$, $v_{1}$ and $v_{2}$ energy bands, respectively. Each LL group is
further split into two spin-dependent LL subgroups because the cooperation
of SOC and magnetic field destroys the spin degeneracy. For any ($%
k_{x},k_{y} $) states, these LLs are doubly degenerate owing to the mirror
symmetry about the z axis. For example, at ($k_{x}=0,k_{y}=0$), the two
degenerate wavefunctions have equivalent spatial distributions, but are
localized near the 0 and 1/2 positions of the enlarged unit cell,
respectively. The 1/2-localization-center states are chosen to illustrate
the main features of LL wavefunctions (Figs. 3(b) and 4(b)). Each LL state
is characterized by the spatial probability density on the A and B
sublattices with sp$^{3}$ orbitals and two spin configurations. This
distribution might have a normal zero-point number and present the
symmetric/anti-symmetric mode about the localization center, as revealed in
a harmonic oscillator (2D electron gas). As a result of the hexagonal
lattice, the A and B sublattices possess the same oscillation mode after
magnetic quantization.

For the first group, the 6p$_{x}$- and 6p$_{y}$-decomposed probability
distributions present an identical oscillation mode in the A$_{\uparrow }$, B%
$_{\uparrow }$, A$_{\downarrow }$ and B$_{\downarrow }$ sublattices (red and
green curves in Fig. 3(b)). Furthermore, they dominate the oscillation modes
of the spin-split LLs, in which the number of zero points can serve as a
quantum number (n$_{\uparrow \downarrow }^{1}$). The n$_{\uparrow }^{1}$ and
n$_{\downarrow }^{1}$ LLs, respectively, have the $\uparrow $- and $%
\downarrow $-dominated components. The splitting of the n$_{\downarrow }^{1}$
and n$_{\uparrow }^{1}$ subgroups are, respectively, shown in Fig. 3(a) by
the solid and dashed blue lines. At ${B_{z}=30}$ T, the first three
conduction LLs belong to the n$_{\downarrow }^{1}$ LLs and then the n$%
_{\uparrow }^{1}=n$ and n$_{\downarrow }^{1}=n+3$ LLs appear alternatively.
The energy spacing between two neighboring LLs in the same subgroup is
almost uniform. This directly reflects the magnetic quantization of the
parabolic $c_{1}$ band (Fig. 2), as observed in 2D electron gas.
Furthermore, the spin-split energy between two different subgroups is about $%
37$ meV regardless of the state energy. The third group is very different
from the first group in terms of energy spacing, split energy; orbital and
spin components. The LL spacing is non-uniform because of the valley-like
energy dispersion. The spin-split energy declines in the increase of state
energy, as shown by the dashed and solid red lines in Fig. 3(a). The
dominating orbitals cover 6p$_{z}$, 6p$_{x}$ and 6p$_{y}$ (black, red and
green curves in the lower half part of Fig. 3(b)). They have the same
quantum mode, while the former and the latter two present the opposite spin
configurations. The 6p$_{z}$ or 6p$_{x}$/6p$_{y}$ component is suitable for
serving as the dominating quantum mode ($n^{3}$), and the latter is chosen
to illustrate the diversified properties among three groups of LLs. For the $%
n_{\uparrow }^{3}$ [$n_{\downarrow }^{3}$] LLs, there are, respectively. $%
n^{3}$ and $n^{3}+1$ [$n^{3}-1$ and $n^{3}$] zero points in the 6p$_{x}$/6p$%
_{y}$-dependent (A$^{\uparrow }$,B$^{\uparrow }$) and (A$^{\downarrow }$,B$%
^{\downarrow }$) sublattices. The similar zero-point numbers are revealed in
the ${n_{\downarrow }^{2}}$ LLs (Fig. 4(b)).

Specially, the second group of LLs exhibits an abnormal ordering (Fig. 4(a))
and the highly asymmetric probability distributions (Fig. 4(b)). The quantum
number n$_{\uparrow \downarrow }^{2}$ is determined from the 6p$_{x}$/6p$%
_{y} $-dependent oscillation mode even if its probability density is lower
than that of the 6p$_{z}$ orbital. The spatial probability density does not
present the well-behaved symmetric distribution about the localization
center except for the n$_{\uparrow \downarrow }^{2}=0$ and/or 1 LLs. This
suggests the superposition of the main and side oscillation modes (the
distinct normal modes) in each n$_{\uparrow \downarrow }^{2}$ LL.
Apparently, the LL energy spacing and the spin-split energy do not have the
specific relations with state energy, as shown by the dashed (n$_{\uparrow
}^{2}$) and solid (n$_{\downarrow }^{2}$) black lines in Fig. 4(a). The
unusual LL energy spectrum, being sensitive to the magnetic-field strength,
is related to the sombrero-shaped energy dispersion (Fig. 2). At ${B_{z}=30}$
T, the $\uparrow $-dominated ($\downarrow $-dominated) LL energies have the
ordering of ${E^{v}(n_{\uparrow }^{2})}$${>E^{v}(n_{\uparrow }^{2}-1)}$ for n%
$_{\uparrow }^{2}{\leq \,4}$ (n$_{\downarrow }^{2}{\leq \,5}$), and then the
inverse ordering for others. This clearly reflects the small-n$_{\uparrow
\downarrow }^{2}$ LLs arising from the inner valley centered at the $\Gamma $
point (Fig. 2), the n$_{\uparrow \downarrow }^{2}$-dependent energy ordering
similar to the wave-vector dependence of energy band, and the strong
competitions between the inner and outer constant-energy loops.

The low-lying LLs exhibit the rich and unique B$_{z}$-dependent energy
spectra, as clearly illustrated in Figs. 5 and 6(a). The conduction LLs has
no intra-subgroup and inter-subgroup crossings/anti-crossings (the dashed
and solid blue curves in Fig. 5(a)). Each LL energy exhibits the linear $%
B_{z}$-dependence, in which the neighboring LL spacing and the spin-split
energy are proportional to the field strength. These are directly reflected
from the monotonous wave-vector dependence of a parabolic conduction band.
The third- and second-group LLs coexist in the deeper valence energy
spectrum (the red and black curves in Fig. 5(b)). They frequently cross each
other because of the well-behaved spatial distributions without the same
quantum mode. The former roughly have the $\sqrt{B_{z}}$-dependent energies,
especially for the larger n$_{\uparrow \downarrow }^{3}$ or the stronger $%
B_{z}$. This is associated with the quasi-linear valence band.

The LL energy spectrum of the second group, as shown in Fig. 6(a), is in
sharp contrast with those of the first and third groups (Figs. 5(a) and
5(b)). All the LLs present the arc-like $B_{z}$-dependence except that the {n%
}${_{\downarrow }^{2}=0}$ LL energy monotonously declines with the
increasing $B_{z}$. Their energies agree with that of the $\Gamma $ point ($%
-0.222$ eV) in the sombrero-shaped band (Fig. 2(a)) when $B_{z}$ approaches
zero. This clearly indicates that the magnetic quantization is initiated
from electronic states near the $\Gamma $ point. For very large n$_{\uparrow
\downarrow }^{2}$ (${>30}$), the LL energies grow quickly as $B_{z}$
slightly increases from zero. The {n}${{_{\uparrow }^{2}}}$ and {n}${%
_{\downarrow }^{2}}$ LLs reach the maximum energy ($-0.146$ eV), being
nearly identical at ${B_{z}\sim \,1}$ T. This energy corresponds to the
highest level of the outer constant-energy loop. With the increase of $B_{z}$%
, the spin-dependent two subgroups start to separate, in which the {n}${{%
_{\uparrow }^{2}}}$ LLs exhibit more drastic changes. As a result, there
exist very frequent inter-subgroup crossings and anti-crossings, depending
on whether the neighboring {n}${{_{\uparrow }^{2}}}$ and {n}${_{\downarrow
}^{2}}$ LLs have the same oscillation modes.

The anti-crossings between two spin-dominated subgroups deserve a closer
examination. They mainly arise from the {n}${{_{\uparrow }^{2}}=n}$ and n$%
_{\downarrow }^{2}{\,=n+4}$ LLs, as illustrated by the red rectangles in
Fig. 6(a). In addition to the major mode with $n$ zero points, these two LLs
also possess the side modes with different zero-point numbers. The latter
are due to the cooperation of the intrinsic interactions and the magnetic
field. Such modes are examined to have ${n\pm \,3}$ zero points by the
detailed numerical calculations. For example, the n$_{\downarrow }^{2}{\,=6}$
LL strongly anti-crosses with the {n}${{_{\uparrow }^{2}}\,=2}$ LL in the
range of 60 T${<B_{z}<70}$ T (Fig. 6(b)). When $B_{z}$ increases from 50 T
along the higher-energy path (the solid blue arrow), the ${n=6}$ main mode
in the A$^{\downarrow }$/B$^{\downarrow }$ sublattice declines quickly (Fig.
6(c)), and the n = 3 side mode in the same sublattice grows rapidly.
Furthermore, the main and side modes, respectively, with 5 and 2 zero points
in the A$^{\uparrow }$/B$^{\uparrow }$ sublattice behave similarly. These
two modes are comparable near the critical magnetic field ($\sim $65 T).
Their roles are interchanged in the further increase of field strength. For
example, at ${B_{z}=80}$ T, the spatial distribution is dominated by the
oscillation mode with 2 zero points in the A$^{\uparrow }$ and B$^{\uparrow
} $ sublattices (the first row in Fig. 6(c)). That is, the ${n_{\downarrow
}^{2}\,=6}$ LL is changed into the ${n_{\uparrow }^{2}\,=2}$ LL during the
variation of $B_{z}$. The probability transfer between the spin-up and
spin-down components is driven by the critical SOC. The similar
anti-crossing process is revealed in the inverse transformation along the
lower-energy path (the dashed blue arrow).

The van Hove singularities in the energy-wave-vector space can create the
special structures in DOS, being sensitive to the effective dimensions. DOS
is defined as
\begin{equation}
D(E)=\sum\limits_{n_{\uparrow \downarrow }^{c,v}}\int\nolimits_{1st\text{ }%
Bz}\frac{\Gamma ^{\prime }}{[E^{c,v}(n_{\uparrow \downarrow
}^{c,v},k_{x},k_{y})-E]^{2}+\Gamma ^{\prime 2}}dk_{x}dk_{y}.
\end{equation}%
${\Gamma ^{\prime }}$ (=0.1 meV) in the calculations is the broadening
parameter. At zero field, the 2D band structure exhibits three shoulder
structures (the red dashed circles) and one strong asymmetric peak in the
square-root form (the red solid circle), as shown in Figs. 7(a) and 7(b).
The first, second and third shoulders situated at ${E\,=0.148}$ eV, ${-0.222}
$ eV and ${-0.396}$ eV are, respectively due to the band-edge states (the
extreme points) in the parabolic conduction band, the rounded inner
constant-energy loop centered at the $\Gamma $ point, and the rounded
valley-like valence band (Fig. 2(a)). The latter comes from the outer
constant-energy loop with the highest level in the first valence band, since
it could be regarded as a 1D parabolic band. Band gap is energy difference
between the first shoulder of the conduction states (Fig. 7(a)) and the
prominent asymmetric peak. Under the magnetic quantization, the
delta-function-like peaks arising from the zero-dimensional LLs come to
exist. Their intensities are proportional to the number of LLs. The first
group, as shown by the black solid curves in Fig. 7(a) at ${B_{z}=30}$ T,
presents a lot of uniform symmetric peaks, being composed of the ${\uparrow }
$- and ${\downarrow }$-dominated ones. This further illustrates the absence
of crossing and anti-crossing. Specially, the initial three conduction peaks
nearest to $E_{F}$ are associated with the ${n_{\downarrow }^{1}\,=0-2}$ LLs
(the blue circles), indicating the specific energy spacing between two
neighboring LLs. For the higher-energy peaks, they could be utilized to
identify the separate contributions of the n$_{\uparrow \downarrow }^{1}$
LLs. As to the valence LLs, they exhibit many double-peak structures and
some single peaks (Fig. 7(b)). The former are induced by the frequent
crossings and anti-crossings between the ${n_{\uparrow }^{2}}$ and ${%
n_{\downarrow }^{2}}$ LLs, and the similar crossings between the n$%
_{\uparrow \downarrow }^{2}$ and n$_{\uparrow \downarrow }^{3}$ LLs. The
higher-energy valence peaks before the strong asymmetric peak are due to the
${n_{\downarrow }^{2}}$ LLs. Their energies and numbers could be tuned by
the magnetic field strength. The threshold peak energy will, respectively,
approach to $-0.148$ eV and $-0.222$ eV (energies of the asymmetric peak and
the second shoulder), when ${B_{z}}$ is reduced to 1 T and $\sim 0$ T.

The diverse magnetic quantization phenomena of monolayer bismuthene are
explored in detail using the generalized tight-binding model. The main
features of LLs are determined by the multi-orbital chemical bondings, the
distinct site energies, the nearest and next-nearest hopping integrals, the
significant spin-orbital couplings and the magnetic field. The theoretical
model could be further developed to solve new Hamiltonians in emerging 2D
materials under the uniform/non-uniform external fields, e.g., the magnetic
Hamiltonians with various interactions in few-layer bismuthene, antimonene
\cite{Sb1,Sb2}, phosphorene \cite{P1,P3} and arsenene \cite{Bigeo1,As,As1}.
Moreover, the generalized tight-binding model could combine with the single-
and many-particle theories to study the other essential physical properties,
such as magneto-optical spectra \cite{optical}, magnetoplasmons \cite%
{plasmon}, and quantum transports \cite{transport}.

The low-energy electronic structure covers the parabolic conduction band,
the sombrero-shaped valence band (the inner and outer constant-energy loops)
and the rounded valley-like valence band, with an indirect gap of $0.293$
eV. Such energy bands are closely related to three groups of (6p$_{x}$,6p$%
_{y}$,6p$_{z}$)--created LLs, in which each group is split into the $%
\uparrow $- and $\downarrow $-dominated subgroups. The dominating
oscillation mode of the 6p$_{x}$/6p$_{y}$-projected probability distribution
could provide a good quantum number. The first and third groups possess the
well-behaved wave functions, while the second group might have the main and
side modes in the n$_{\uparrow \downarrow }^{2}$ LLs. The former two do not
present the anti-crossing behavior in the ${B_{z}}$-dependent energy
spectra. However, the frequent anti-crossings occur between the ${%
n_{\uparrow }^{2}=n}$ and ${n_{\downarrow }^{2}=n+4}$ LLs. The LL energies
of the first, second and third groups, respectively, exhibit the linear,
arc-shaped and square-root-form ${B_{z}}$-dependences. Furthermore, the
normal spin-split energies are only revealed in the first group. The van
Hove singularities in parabolic bands, the outer constant-energy loop and
LLs, respectively, lead to shoulders, a prominent asymmetric peak and many
delta-function-like peaks. The predicted electronic energy spectra could be
examined by STS measurements.

\noindent \textit{Acknowledgments.} This work was supported by the MOST of Taiwan, under Grant No. MOST 105-2112-M-006-002-MY3 and MOST 106-2112-M-022-001.

\noindent ~~~~$^\ast$e-mail address:szuchaochen@gmail.com;$^\star$e-mail
address:yarst5@gmail.com;$^\dagger$e-mail address: mflin@mail.ncku.edu.tw.

\bigskip \vskip0.6 truecm

\noindent

\newpage

\textbf{Figure captions}

Figure 1: (a) The $x-y$ projections of monolayer bismuthene with an enlarged
rectangular unit cell in ${B_{z}}$ ${\hat{z}}$, (b) the buckled structure,
(c) the various orbital hybridizations; (d) the Brillouin zone of the
hexagonal lattice with the high-symmetry points and the reduced first
Brillouin zone. $\mathbf{a_{1}}$ and $\mathbf{a_{2}}$ in (a) are lattice
vectors, and the subscript of A$_{i}$ corresponds to the $i$-th Bi atom.

Figure 2: Energy bands of bismuth along the high-symmetry points. Also shown
are the 3D band structures near the $\Gamma $ point.

Figure 3: (a) The spin-dependent LL energies for the first and the third
groups at ${B_{z}=30}$ T; (b) the orbital-projected probability
distributions in the A$^{\uparrow }$, B$^{\uparrow }$, A$^{\downarrow }$ and
B$^{\downarrow }$ sublattices.

Figure 4: Similar plot as Fig. 3, but shown for the second group of LLs.

Figure 5: The $B_z$-dependent LL energy spectra for the first, second and
third groups (the blue, black and red curves) in the absence/presence of
crossings.

Figure 6: (a) The crossing and anti-crossing energy spectra due to the
spin-up- and spin-down-dominated second group of LLs, (b) the ${%
n_{\downarrow }^{2}\,=6}$ and ${n_{\uparrow }^{2}\,=2}$ LL anti-crossing
within a certain range of $B_{z}$; the variations of probability
distributions along the (c) higher- and (d) lower-energy paths.

Figure 7: The low-energy density of states of monolayer bismuthene under ${%
B_z=0}$ and 30 T for (a) conduction and (b) valence states.

\begin{figure}[p]
\centering
\includegraphics[width=0.75\textwidth]{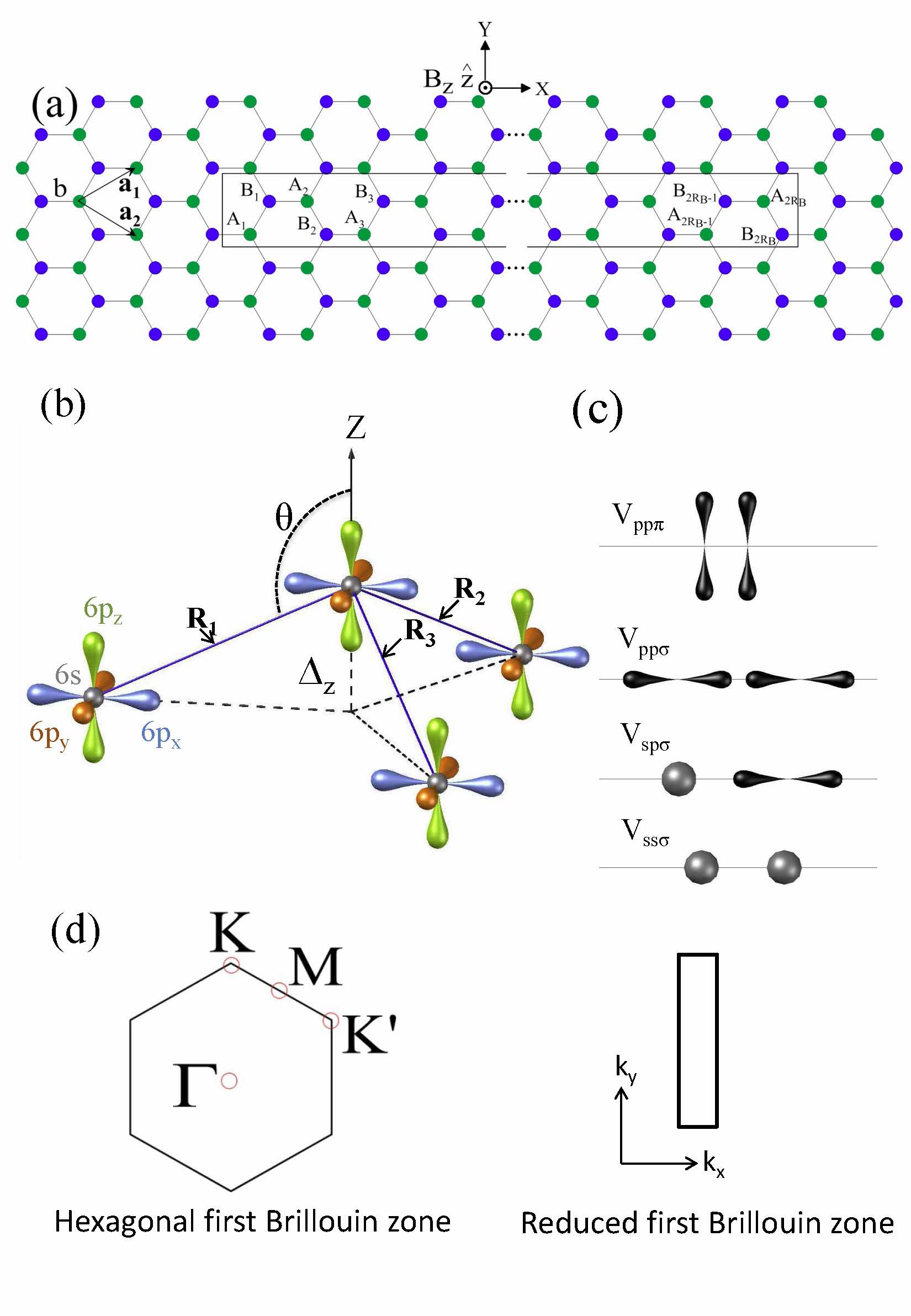}
\caption{(a) The $x-y$ projections of monolayer bismuthene with an enlarged
rectangular unit cell in ${B_{z}}$ ${\hat{z}}$, (b) the buckled structure,
(c) the various orbital hybridizations; (d) the Brillouin zone of the
hexagonal lattice with the high-symmetry points and the reduced first
Brillouin zone. $\mathbf{a_{1}}$ and $\mathbf{a_{2}}$ in (a) are lattice
vectors, and the subscript of A$_{i}$ corresponds to the $i$-th Bi atom.}
\label{figure:1}
\end{figure}

\begin{figure}[p]
\centering
\includegraphics[width=0.85\textwidth]{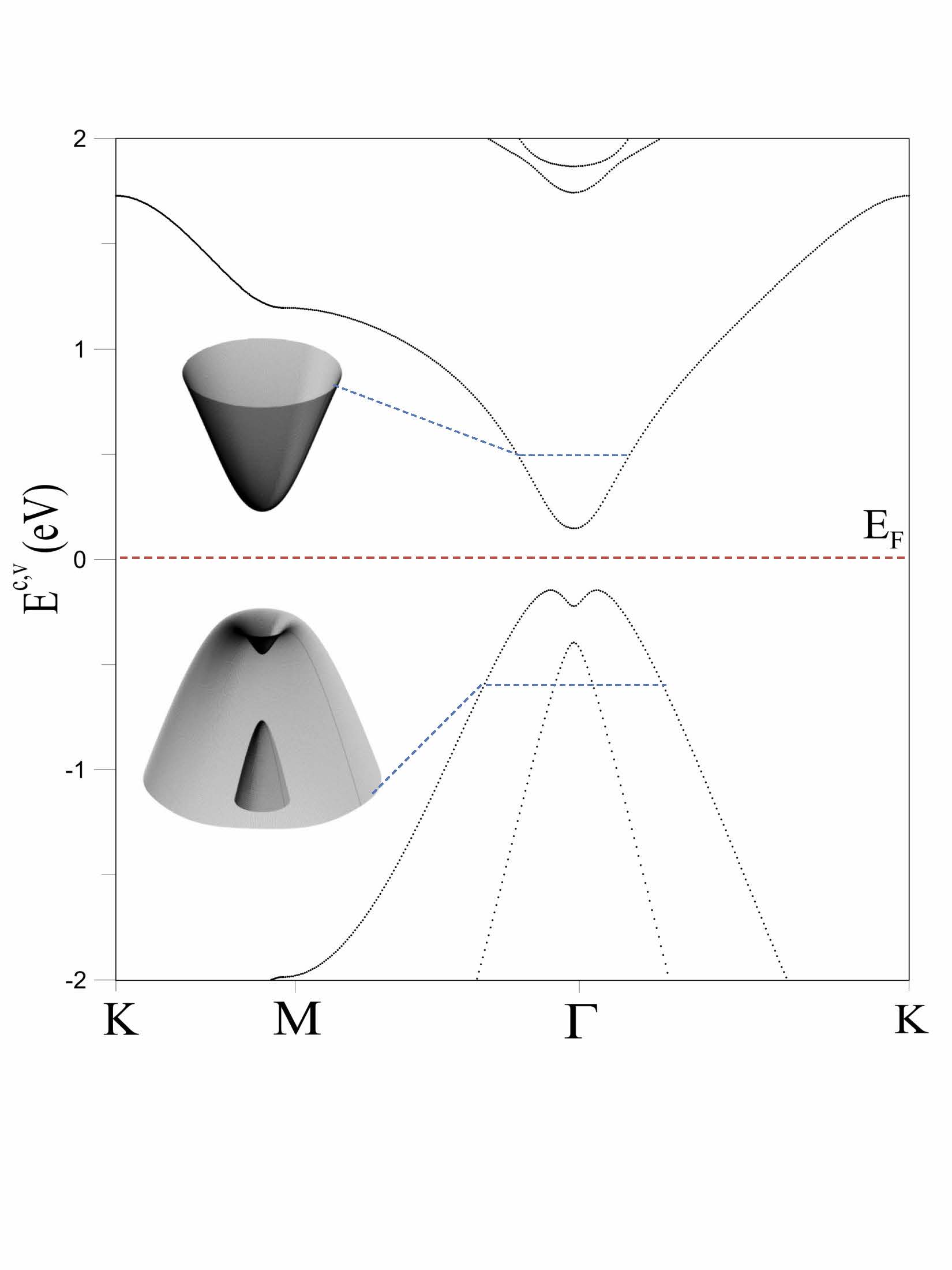}
\caption{Energy bands of bismuth along the high-symmetry points. Also shown
are the 3D band structures near the $\Gamma $ point.}
\label{figure:2}
\end{figure}

\begin{figure}[p]
\centering
\includegraphics[width=0.85\textwidth]{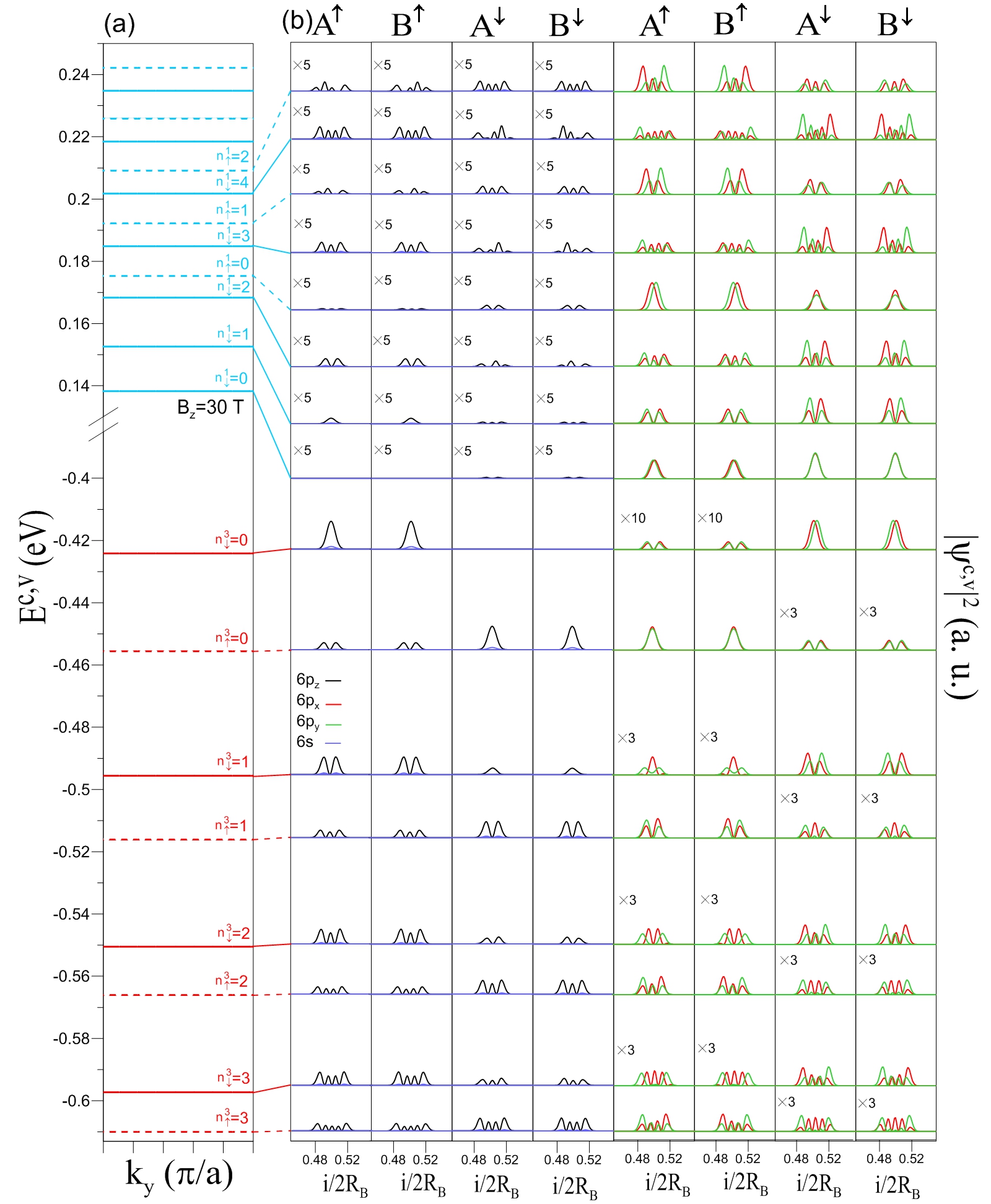}
\caption{(a) The spin-dependent LL energies for the first and the third
groups at ${B_{z}=30}$ T; (b) the orbital-projected probability
distributions in the A$^{\uparrow }$, B$^{\uparrow }$, A$^{\downarrow }$ and
B$^{\downarrow }$ sublattices.}
\label{figure:3}
\end{figure}

\begin{figure}[p]
\centering
\includegraphics[width=0.85\textwidth]{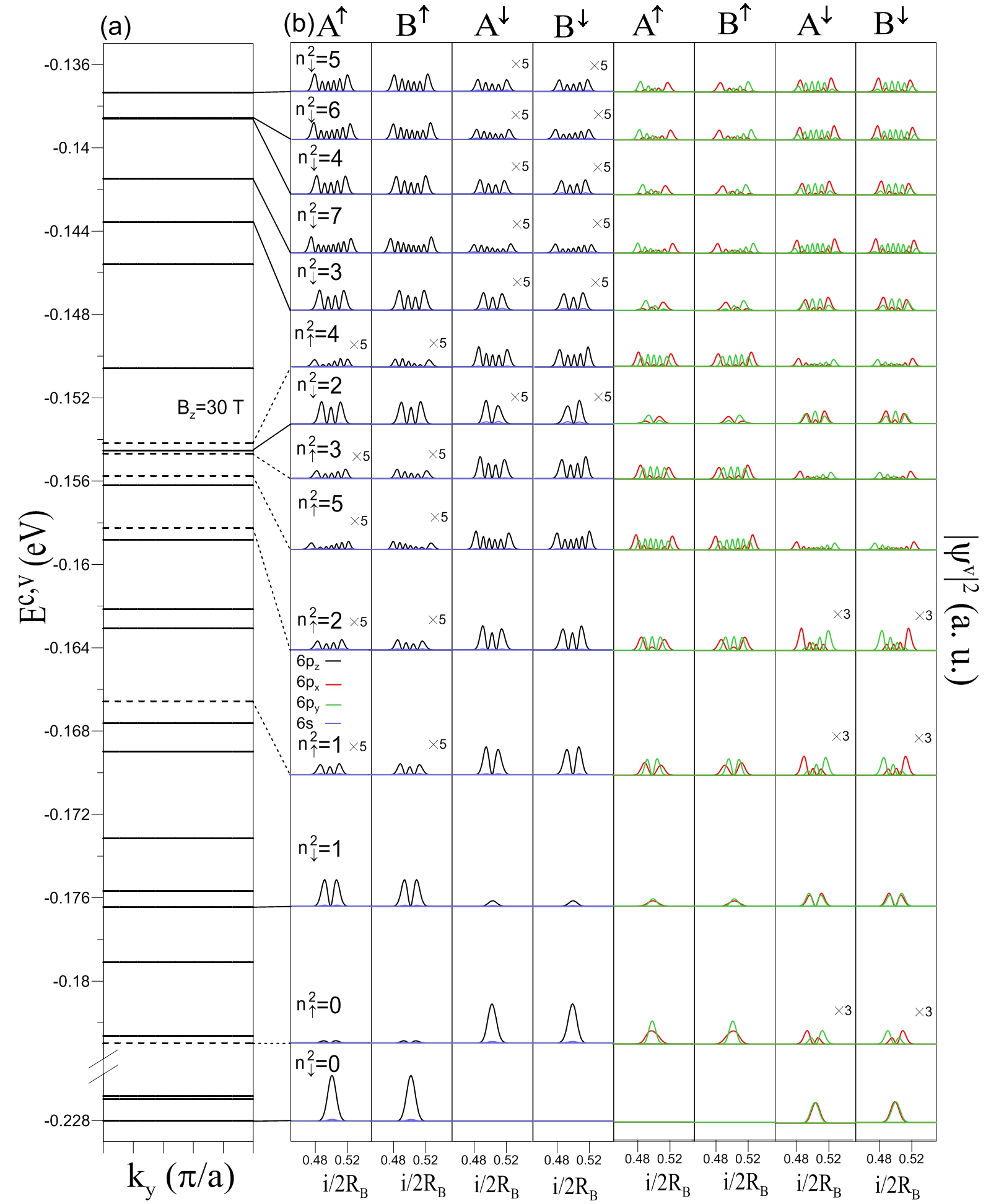}
\caption{Similar plot as Fig. 3, but shown for the second group of LLs.}
\label{figure:4}
\end{figure}

\begin{figure}[p]
\centering
\includegraphics[width=0.65\textwidth]{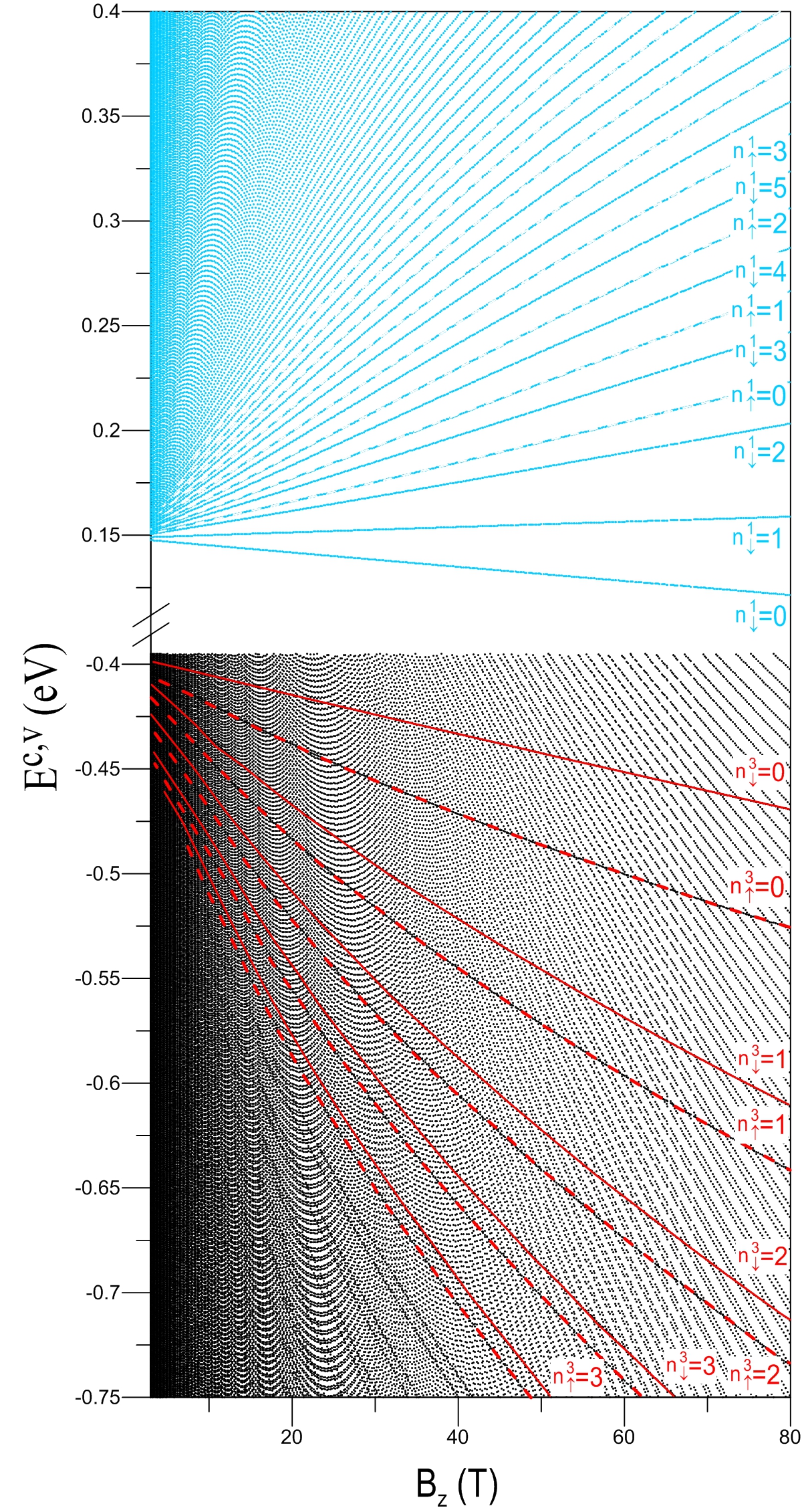}
\caption{The $B_z$-dependent LL energy spectra for the first, second and
third groups (the blue, black and red curves) in the absence/presence of
crossings.}
\label{figure:5}
\end{figure}

\begin{figure}[p]
\centering
\includegraphics[width=0.85\textwidth]{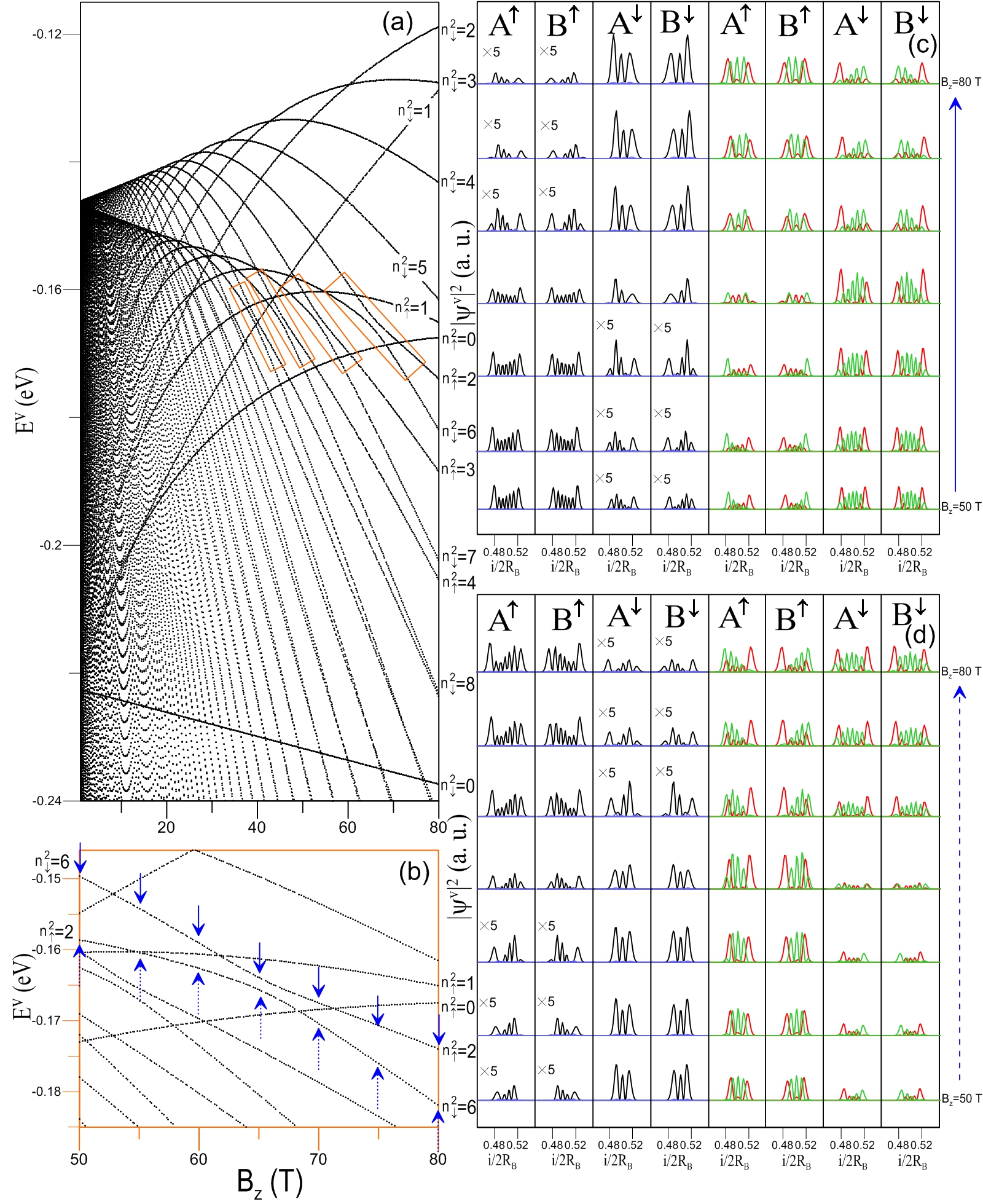}
\caption{(a) The crossing and anti-crossing energy spectra due to the
spin-up- and spin-down-dominated second group of LLs, (b) the ${%
n_{\downarrow }^{2}\,=6}$ and ${n_{\uparrow }^{2}\,=2}$ LL anti-crossing
within a certain range of $B_{z}$; the variations of probability
distributions along the (c) higher- and (d) lower-energy paths.}
\label{figure:6}
\end{figure}

\begin{figure}[p]
\centering
\includegraphics[width=0.85\textwidth]{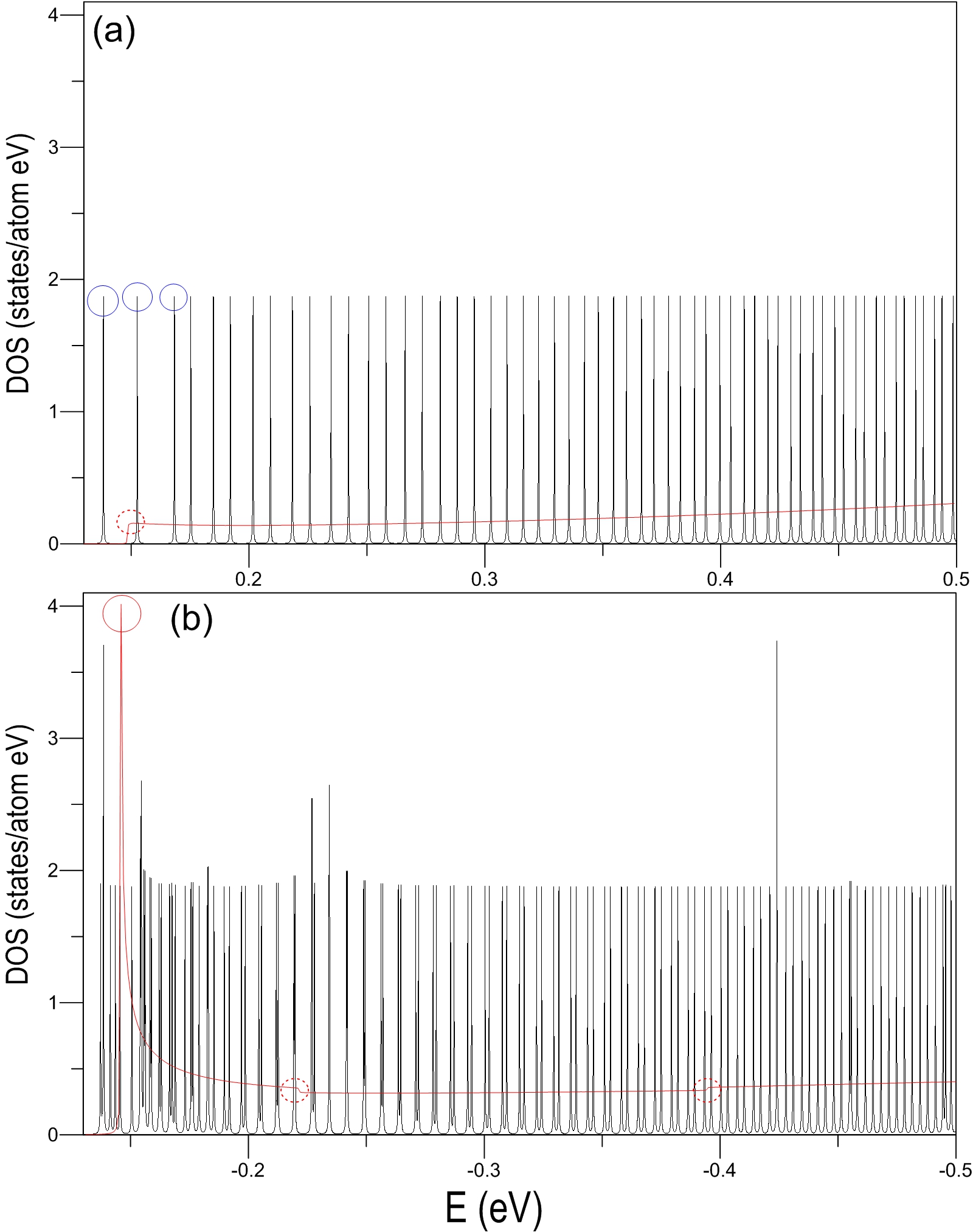}
\caption{The low-energy density of states of monolayer bismuthene under ${%
B_z=0}$ and 30 T for (a) conduction and (b) valence states.}
\label{figure:7}
\end{figure}

\end{document}